\documentclass{aa}

\usepackage{graphicx}
\usepackage{txfonts}

\usepackage{xcolor}
\usepackage{xspace}
\usepackage[colorlinks,urlcolor={blue!80!black},citecolor={blue!60!black},linkcolor={red!60!black}]{hyperref}
\usepackage{multirow}
\usepackage{caption}
\usepackage{subcaption}

\newcommand{\frbpoppy}{\texttt{frbpoppy}\xspace}
\newcommand{\lognlogs}{$\log N$-$\log S$\xspace}
\def\pop#1{\texttt{#1}\xspace}
\def\survey#1{\texttt{#1}\xspace}
\def\est#1{\textcolor{gray}{#1}\xspace}
\newcommand{\paperone}{2019A&A...632A.125G}
\newcommand{\package}[2]{\texttt{#1} \citep{#2}}

\makeatletter
\renewcommand*\aa@pageof{, page \thepage{} of \pageref*{LastPage}}
\makeatother
\begin{document}
\title{Multi-dimensional population modelling using \frbpoppy: \\ magnetars can produce the observed Fast Radio Burst sky}
\titlerunning{Multi-dimensional population modelling using \frbpoppy:}

\author{D.W. Gardenier
 \inst{1, 2}
 \and
 J. van Leeuwen\inst{1, 2}\fnmsep\thanks{\email{leeuwen@astron.nl}}
}
\institute{ASTRON - the Netherlands Institute for Radio Astronomy, Oude Hoogeveensedijk 4, 7991 PD, Dwingeloo, The Netherlands
 \and
 Anton Pannekoek Institute for Astronomy, University of Amsterdam, Science Park 904, 1098 XH Amsterdam, The Netherlands
}

\date{Received December 11, 2020; accepted Apr 21, 2021}

\abstract{
Fast Radio Bursts (FRBs) are energetic, short, bright  transients that occur frequently over the entire radio sky.
The observational challenges following from their fleeting, generally one-off nature have prevented identification of the underlying sources producing the bursts.
As the population of detected FRBs grows, the observed distributions of brightness, pulse width and dispersion measure now begin to take shape.
Meaningful direct interpretation of these distributions is, however, made impossible by the selection effects that telescope and search pipelines  invariably imprint on each FRB survey.
Here we show that multi-dimensional FRB population synthesis can find a single, self-consistent population of FRB sources that can reproduce the real-life results of the major ongoing FRB surveys.
This means that individual observed distributions can now be combined to derive the properties of the intrinsic FRB source population.
The characteristics of our best-fit model for one-off FRBs agree with a population of magnetars.
We extrapolate this model and predict the number of FRBs future surveys will find.
For surveys that have commenced, the method we present here can already determine the composition of the FRB source class, and potentially even its subpopulations.}

\maketitle

\section{Introduction}
Fast Radio Bursts (FRBs) are bright millisecond-duration radio transients of unknown origin \citep[see][]{petroff2019review, cordes2019review}.
Recent observations of SGR~1935+2154 suggest
that some FRBs can be associated with magnetars \citep{2020Natur.587...59B, chime2020}.
Despite these detections, much is still unknown about the intrinsic FRB source population.
The number density function could be fairly flat \citep{2017AJ....154..117L} or quite steep \citep{2019MNRAS.483.1342J}.
The luminosity function could be described by a power law distribution \citep{2016MNRAS.458..708C} or a Schechter function \citep{2018MNRAS.481.2320L, 2018ApJ...863..132F}.
The difference in intrinsic pulse width distribution of repeating and one-off FRB sources could be due to an intrinsic difference \citep{2019ApJ...885L..24C, fonseca2020, 2020MNRAS.tmp.3174C} or due to selection effects \citep{2020MNRAS.497.3076C}.

In recent years a number of observatories have focused on detecting larger numbers of FRBs and studying these in more
detail, to help solve some of the open questions listed above.  These  telescopes and surveys include  the Australian Square Kilometre Array Pathfinder  \citep[ASKAP;][]{craft, askap}, the FRB survey on Canadian Hydrogen Intensity Mapping Experiment  \citep[CHIME/FRB;][]{chimeoverview}, and the Apertif survey at Westerbork \citep{arts, lk+21}.
By now, over a hundred FRB detections have been made \citep{tns}.
This order of magnitude increase in detected FRBs provides a unique opportunity to conduct population studies.

Prior FRB population studies have focussed on different aspects, such as \citet{2016MNRAS.458..708C} who conducted one of the first population studies, investigating whether FRB sources could be considered to have a cosmological, high redshift, origin.
Just a selection of these show papers focused on
pulse broadening \citep{2020MNRAS.497.1382Q},
spectral properties \citep{2019ApJ...872L..19M},
source count distributions \citep{2019MNRAS.483.1342J},
fluence and dispersion measure distributions \citep{2019ApJ...883...40L}
or a mix of properties from scattering to pulse width \citep{2020ApJ...899..124B}.
Few of these earlier studies incorporate all three  aspects we consider important when conducting a full scale population study: selection effects, describing detections from multiple surveys and making the code available.
Extrapolating from observations to the properties of an intrinsic population is difficult without taking the full range of selection effects for different surveys into account.

Population synthesis offers the unique opportunity to probe an intrinsic source population through extensive modelling.
By simulating an intrinsic source population and convolving this population with a model of selection effects, a simulated observation can be made.
Comparing this modelled observation to real observations provides a means to evaluate the simulated intrinsic population.
The better the simulated observed population fits real observations, the more likely the simulated intrinsic population is to be an accurate descriptor of the real intrinsic population.
This method has been applied successfully, and extensively, in a wide range of fields: from pulsars \citep{taylor77} to high-mass binaries \citep{1996A&A...309..179P}, from gamma ray bursts \citep{GRBs} to stellar evolution  \citep{2018arXiv180806883I}.

In this paper we aim to determine which physical birth and emission properties best describe the intrinsic one-off FRB population.
We do this through a large scale FRB population synthesis, exploring an 11-dimensional parameter-space through Monte Carlo simulations.

This paper comprises of the following sections.
Firstly, our methods in Sect.~\ref{sec:methods}.
Secondly, our results in Sect.~\ref{sec:results}, split into three parts.
We start by showing the effect of various intrinsic parameters on a \lognlogs distribution.
We then show the results of a Monte Carlo simulation through which we derive the optimal intrinsic parameters with which the observed one-off FRB population can be described.
We subsequently use these optimal parameters to predict expected detections by future radio transient surveys.
Thirdly, a discussion of the results in Sect.~\ref{sec:discussion} and
lastly, our conclusions in Sect.~\ref{sec:conclusions}.

\section{Methods}
\label{sec:methods}
One way to probe an intrinsic source population is through population synthesis.
An advantage of population synthesis is that it takes into account both the physics of the underlying population, and
selection effects.
Synthesizing populations requires an intrinsic source population to be modelled before convolving it with a range of selection effects to obtain an observed population.
Comparing such a population to real observed populations subsequently allows various intrinsic source population
parameters to be evaluated, and determined.
The better the simulation, the better the inputs are able to describe the true intrinsic population.

In \citet{2019A&A...632A.125G} we presented \texttt{v1.0} of \frbpoppy, an open source code package capable of Fast Radio Burst population synthesis in Python.
We subsequently expanded its functionality to include the modelling of repeating FRB sources in \texttt{v2.0} of \frbpoppy \citep{2020arXiv201202460G}.
We build on this prior work in this paper, and have accordingly published the code used for these results in \texttt{v2.1} of \frbpoppy\footnote{See \url{https://github.com/davidgardenier/frbpoppy}}.

In this paper, we aim to provide a method by which intrinsic FRB source population parameters can be derived.
We do so by adopting an iterative approach to population synthesis, using the outcome of a simulation to determine the choice of input parameters for subsequent runs.
This allows for the properties of an intrinsic source population which best describe current observations to be derived.

We use \texttt{v2.0} of \frbpoppy to model our intrinsic populations, surveys and observed populations.
Modelling these components requires various parameters as input.
For the modelling of intrinsic populations we adopt the parameters given in Table~\ref{tab:populations} and for surveys we adopt the parameters given in Table~\ref{tab:surveys}.
A number of these parameters may require further explanation.
Specifically, in \frbpoppy, the luminosity of a burst refers to the isotropic equivalent bolometric luminosity in the radio, where
the frequency range is defined by $\nu_{\text{low, high}}$.
When drawing such luminosities from a power law, we define a power law to follow $N(L)\propto L^{{ \text{li} }}$ with luminosity index ${ \text{li} }$.
Additionally, log-normal distributions in \frbpoppy use the desired mean and standard deviation given as input to calculate the mean and standard of said variables natural logarithm. These values subsequently allow a log-normal distribution to be drawn around the desired mean and with the desired spread.
For additional information on the modelling process, or the parameter inputs we refer the reader to
\citet{2019A&A...632A.125G, 2020arXiv201202460G}.

\begin{table*}
 \caption{
  The parameters and values used to model intrinsic FRB populations in this paper.
  Listed are the number of generated sources $n_{\rm gen}$, maximum timescale in terms of number of days $n_{\rm days}$ and whether generating a repeater population `repeaters'.
  Number density parameters $\rho$ include the number density model $n_{\rm model}$ and cosmological parameters, Hubble constant $H_0$, density parameter $\Omega_{\rm m}$, cosmological energy density due to the cosmological constant $\Omega_{\Lambda}$, number density parameter $\alpha$, as well as the maximum redshift $z_{\text{max}}$.
  Dispersion measure (DM) components include contribution from the host DM$_{\rm host}$, from the intergalactic medium DM$_{\rm igm}$ and from the Milky Way DM$_{\rm mw}$, each with a particular model and related parameters.
  DM$_{\rm tot}$ reflects whether particular DM components are modelled or not.
  Furthermore there is the emission range $\nu_{\textrm{emission}}$, the isotropic equivalent bolometric luminosity in radio L$_{\rm bol}$, spectral index $\text{si}$ and intrinsic pulse width $w_{\text{int}}$, all with their respective modelling parameters.
  A empty fields indicates a particular argument was not required for the generation of that population.
  \label{tab:populations}
 }
 \centering
\begin{tabular}{c c c c c c c}
 \hline\hline                                                                                                                              \\[-9px]
 Parameters                & Arguments          & Units                  & \pop{simple}        & \pop{complex}       & \pop{optimal}       \\[1px]
 \hline                    &                    &                        &                     &                     &                     \\[-9px]
                           & $n_{\rm gen}$      &                        & $10^{5}$            & $5 \cdot 10^7$      & $5 \cdot 10^7$      \\
                           & $n_{\rm days}$     & days                   & 1                   & 1                   & 1                   \\
                           & repeaters          &                        & False               & False               & False               \\[1px]
 \hline                    &                    &                        &                     &                     &                     \\[-9px]
 $\rho$                    & $n_{\rm model}$    &                        & vol$_{\textrm{co}}$ & vol$_{\textrm{co}}$ & vol$_{\textrm{co}}$ \\
                           & $\text{H}_{0}$     & km s$^{-1}$ Mpc$^{-1}$ & 67.74               & 67.74               & 67.74               \\
                           & $\Omega_{\rm m}$   &                        & 0.3089              & 0.3089              & 0.3089              \\
                           & $\Omega_{\Lambda}$ &                        & 0.6911              & 0.6911              & 0.6911              \\
                           & $\alpha$           &                        & $-1.5$              & $-1.5$              & $-2.2$                \\
                           & $z_{\rm max}$      &                        & 0.01                & 1                   & 2.5                 \\[1px]
 \hline                    &                    &                        &                     &                     &                     \\[-9px]
 DM$_{\rm host}$           & model              &                        &                     & gauss               & constant            \\
                           & mean               & pc cm$^{-3}$           &                     & 100                 &                     \\
                           & std                & pc cm$^{-3}$           &                     & 200                 &                     \\
                           & value              & pc cm$^{-3}$           &                     &                     & 50                  \\[1px]
 \hline                    &                    &                        &                     &                     &                     \\[-9px]
 DM$_{\rm igm}$            & model              &                        &                     & ioka                & ioka                \\
                           & mean               & pc cm$^{-3}$           &                     &                     &                     \\
                           & std                & pc cm$^{-3}$           &                     & 200                 & 200                 \\
                           & slope              & pc cm$^{-3}$           &                     & 1000                & 1000                \\[1px]
 \hline                    &                    &                        &                     &                     &                     \\[-9px]
 DM$_{\rm mw}$             & model              &                        &                     & ne2001              & ne2001              \\[1px]
 \hline                    &                    &                        &                     &                     &                     \\[-9px]
 DM$_{\rm tot}$            & host               &                        & False               & True                & True                \\
                           & igm                &                        & False               & True                & True                \\
                           & mw                 &                        & False               & True                & True                \\[1px]
 \hline                    &                    &                        &                     &                     &                     \\[-9px]
 $\nu_{\textrm{emission}}$ & low                & MHz                    & $10^7$              & $10^7$              & $10^7$              \\
                           & high               & MHz                    & $10^9$              & $10^9$              & $10^9$              \\[1px]
 \hline                    &                    &                        &                     &                     &                     \\[-9px]
 L$_{\rm bol}$             & model              &                        & constant            & powerlaw            & powerlaw            \\
                           & low                & erg s$^{-1}$           &                     & $10^{40}$           & $10^{40}$           \\
                           & high               & erg s$^{-1}$           &                     & $10^{45}$           & $10^{45}$           \\
                           & power              &                        &                     & 0                   & $-0.8$                \\
                           & value              &                        & $10^{38}$           &                     &                     \\[1px]
 \hline                    &                    &                        &                     &                     &                     \\[-9px]
 $\text{si}$               & model              &                        & constant            & gauss               & constant            \\
                           & mean               &                        &                     & $-1.4$              &                     \\
                           & std                &                        &                     & 1                   &                     \\
                           & value              &                        & 0                   &                     & $-0.4$                \\[1px]
 \hline                    &                    &                        &                     &                     &                     \\[-9px]
 $w_{\rm int}$             & model              &                        & constant            & lognormal           & lognormal           \\
                           & mean               & ms                     &                     & 0.1                 & $3.6 \cdot 10^{-3}$ \\
                           & std                & ms                     &                     & 1                   & 0.6                 \\
                           & value              & ms                     & 10                  &                     &                     \\[1px]
 \hline                    &                    &                        &                     &                     &                     \\[-9px]
\end{tabular}
\end{table*}

\begin{table*}
 \caption{An overview of the parameters adopted for the simulation of surveys.
  Parameters include survey degradation factor $\beta$, telescope gain $G$, pointing time $t_{\text{point}}$, sampling time $t_{\text{samp}}$, receiver temperature $T_{\text{rec}}$, central frequency $\nu_{\text{c}}$, bandwidth BW, channel bandwidth BW$_{\text{ch}}$, number of polarisations $n_{\text{pol}}$, field-of-view FoV, minimum signal-to-noise ratio S/N, observatory latitude $\phi$, observatory longitude $\lambda$, mount type, and then the minimum to maximum right ascension $\alpha$, declination $\delta$, Galactic longitude $l$, and Galactic latitude $b$.
  Greyed out values indicate an estimated or average value, or a value obtained through private communication.
  Note that many future surveys have parameters which are still subject to change.
 }
 \label{tab:surveys}
 \centering
\begin{tabular}{c p{1.2cm} ccccc}
 \hline\hline                                                                                                                                     \\[-9px]
 Parameter            & Units   & \survey{askap-incoh} & \survey{chime-frb}  & \survey{chord}      & \survey{fast-crafts} & \survey{parkes-htru}  \\
 \hline                                                                                                                                           \\[-9px]
 $\beta$              &         & \est{1.2}            & \est{1.2}           & \est{1.2}           & \est{1.2}            & \est{1.2}             \\
 G                    & K/Jy    & \est{0.09898}        & \est{1.4}           & \est{3.33}          & {16.46 }             & 0.69                  \\
 $t_{\textrm{point}}$ & s       & \est{3600}           & \est{360}           & \est{360}           & \est{60}             & \est{270}             \\
 $t_{\textrm{samp}}$  & ms      & {1.265}              & 1                   & \est{1}             & \est{0.196608}       & 0.064                 \\
 $T_{\textrm{rec}}$   & K       & \est{70}             & 50                  & \est{50}            & {20}                 & \est{28}              \\
 $\nu_{\textrm{c}}$   & MHz     & {1320}               & 600                 & {900}               & {1250}               & 1352                  \\
 BW                   & MHz     & {336.0}              & 400                 & {1200}              & {400}                & 340                   \\
 BW$_{\textrm{ch}}$   & MHz     & {1}                  & 0.390625            & \est{0.390625}      & {0.076}              & {0.390625}            \\
 $n_{\textrm{pol}}$   &         & {2}                  & 2                   & {2}                 & \est{2}              & 2                     \\
 FoV                  & deg$^2$ & {20}                 & \est{164.15}        & \est{200}           & \est{0.031}          & 0.5551                \\
 S/N                  &         & \est{8}              & \est{10}            & \est{8}             & \est{8}              & 8                     \\
 $\phi$               & \degr   & \est{$-26.696$}        & 49.3208             & \est{49.3208}       & {25.6529}            & $-32.9980$              \\
 $\lambda$            & \degr   & \est{116.637}        & $-119.624$            & \est{$-$119.6238}     & {106.8566}           & 148.2626              \\
 Mount                &         & azimuthal            & transit             & transit             & transit              & azimuthal             \\
 $\alpha$             & \degr   & \est{0 -- 360}       & 0 -- 360            & {0 -- 360}          & 0 -- 360             & \est{0 -- 360}        \\
 $\delta$             & \degr   & $-90$ -- \est{30}      & \est{$-40.679$} -- 90 & \est{$-40.679$} -- 90 & {$-14$ -- 66}          & \est{$-90$ -- 90}       \\
 $l$                  & \degr   & \est{$-180$ -- 180}    & $-180$ -- 180         & {$-180$ -- 180}       & $-180$ -- 180          & $-120$ -- 30            \\
 $b$                  & \degr   & \est{$-90$ -- 90}      & $-90$ -- 90           & {$-90$ -- 90}         & $-90$ -- 90            & $-15$ -- 15             \\
 Reference            &         & 1                    & 2                   & 3                   & 4                    & 5                     \\
 \hline                                                                                                                                           \\
 \hline\hline                                                                                                                                     \\[-9px]
 Parameter            & Units   & \survey{perfect}     & \survey{puma-full}  & \survey{ska1-low}   & \survey{ska1-mid}    & \survey{wsrt-apertif} \\
 \hline                                                                                                                                           \\[-9px]
 $\beta$              &         & \est{1.2}            & \est{1.2}           & \est{1.2}           & \est{1.2}            & \est{1.2}             \\
 G                    & K/Jy    & \est{10$^5$}         & \est{166.5}         & \est{10.71}         & \est{17.65}          & \est{0.7}             \\
 $t_{\textrm{point}}$ & s       & \est{86400}          & \est{600}           & \est{600}           & \est{600}            & \est{10800}           \\
 $t_{\textrm{samp}}$  & ms      & \est{0.001}          & \est{1.25}          & {0.05}              & {0.05}               & \est{0.04096}         \\
 $T_{\textrm{rec}}$   & K       & \est{0.01}           & \est{30}            & {30}                & {30}                 & 70                    \\
 $\nu_{\textrm{c}}$   & MHz     & \est{1000}           & {650}               & {150}               & {7175}               & 1370                  \\
 BW                   & MHz     & \est{800}            & {900}               & {200}               & {13650}              & 300                   \\
 BW$_{\textrm{ch}}$   & MHz     & \est{0.001}          & \est{0.390625}      & {0.0054}            & {0.0134}             & 0.1953                \\
 $n_{\textrm{pol}}$   &         & \est{2}              & \est{2}             & {2}                 & {2}                  & 2                     \\
 FoV                  & deg$^2$ & \est{41253}          & \est{32.6}          & {27}                & {0.49}               & 8.7                   \\
 S/N                  &         & \est{10$^{-16}$}     & \est{8}             & \est{8}             & \est{8}              & \est{10}              \\
 $\phi$               & \degr   & \est{0}              & \est{0}             & \est{$-26.7$}         & \est{$-30.72$}         & 52.91474              \\
 $\lambda$            & \degr   & \est{0}              & \est{0}             & \est{116.67}        & \est{21.41}          & 6.603340              \\
 Mount                &         & \est{azimuthal}      & azimuthal           & transit             & azimuthal            & equatorial            \\
 $\alpha$             & \degr   & \est{0 -- 360}       & 0 -- 360            & 0 -- 360            & 0 -- 360             & \est{0 -- 360}        \\
 $\delta$             & \degr   & \est{$-90$ -- 90}      & \est{$-90$ -- 90}     & \est{$-90$ -- 90}     & \est{$-90$ -- 90}      & \est{$-37.1$ -- 90}     \\
 $l$                  & \degr   & \est{$-180$ -- 180}    & $-180$ -- 180         & $-180$ -- 180         & $-180$ -- 180          & \est{$-180$ -- 180}     \\
 $b$                  & \degr   & \est{$-90$ -- 90}      & $-90$ -- 90           & $-90$ -- 90           & $-90$ -- 90            & \est{$-90$ -- 90}       \\
 Reference            &         &                      & 6                   & 7                   & 7                    & 8                     \\
 \hline
\end{tabular}
 \tablebib{(1)~\citet{2018Natur.562..386S}; (2)~\citet{chimeoverview}; (3)~\citet{2019clrp.2020...28V};
   (4)~\citet{2019SCPMA..6259506Z}; (5)~\citet{2010MNRAS.409..619K}; (6)~\citet{2020arXiv200205072C,
     2019BAAS...51g..53S}; (7)~\citet{dewdney2013ska1}; (8)~\citet{oosterloo, arts}.}
\end{table*}

In this first Monte Carlo \frbpoppy paper we limit ourselves to one-off FRB sources.
While including the repeating population could provide stronger constrains on the intrinsic FRB population, modelling
repeater populations would require exponentially more compute resources.
We discuss the implications of this approach in Sect.~\ref{sec:discussion:limitations}.

\subsection{LogN-LogS}
\label{sec:methods:lognlogs}
A \lognlogs distribution is a distribution showing the number of detections of sources above a limiting signal-to-noise (S/N) threshold.
Note that we use \lognlogs to refer to the cumulative distribution of the number of detected FRBs greater than a limiting \emph{S/N ratio}.
While this relates to a similar distribution for \emph{peak flux density}, or indeed  fluence,
a limiting S/N ratio has the advantage of incorporating all survey selection effects without attempting to account for them.
This allows surveys to be compared according to the same metric.

We choose to model such distributions to show the effect of various intrinsic parameters on an observed distribution.
All such simulations start with a \pop{simple} population (see Table~\ref{tab:populations}), intended to model a population with the most basic assumptions.
From this \pop{simple} population, we model populations with varying maximum redshifts $z_{\text{max}}$.
For investigating the effect of different luminosity functions, we additionally generate populations with power laws with different slopes - all within the range of $10^{40} -10^{43}\ \text{ergs s}^{-1}$.
We also simulate populations with different values of the spectral index $\text{si}$, from $-2$ to 2.
Finally, we also investigate the effect of various pulse width models by simulating populations with a value of 10~ms, a normal distribution of values with a mean of 10~ms and standard deviation of 10~ms and lastly a log-normal distribution with an underlying mean of 10~ms and 10~ms.

The difficulty in drawing meaningful conclusions from a single-dimensional view such as the \lognlogs distribution,
described in more detail Sect.~\ref{sec:discussion:lognlogs},
shows the need for a multi-dimensional approach.

\subsection{Monte Carlo}
\label{sec:methods:mc}
By conducting a Monte Carlo simulation we aim to derive properties of the intrinsic FRB population.
Exploring all possible combinations of the parameter space in \frbpoppy would be impractical due to the sheer number of inputs, and computational constraints.
Instead, we choose subsets of parameters over which we iterate, before shifting to a next set.
To ensure our results convergence towards a global maximum, we perform additional runs in which we return to prior parameters sets.
Better fits within that parameter space indicate a better global fit has been found.
That is in essence what a Monte Carlo simulation is --- a method in which more runs results in a better result.
In our case, a better result is when a simulation provides a more accurate representation of an observed population than a previous simulation.
We chose to measure our succes in terms of a total Goodness of Fit (GoF).
To avoid optimising towards a local maximum, we evaluate each simulation on multiple areas.

We assign three `goodness of fit' estimators to each simulated population, reflecting their measure of success.

The first is the $p$-value corresponding to a two sample Kolmogorov–Smirnov (KS) test applied to the
simulated and corresponding real dispersion measure (DM) distribution.
This real distribution is obtained by using the \texttt{frbcat} package \citep{pipfrbcat} to access the FRB catalogue hosted on the Transient Name Server (TNS; \citealt{tns}).
We choose to filter out any repeating FRB sources to avoid adding significant weight to their observed parameters.
By grouping the database by survey, we allow a separate DM and S/N distribution to be derived for each survey.
The results in this paper are based on the TNS catalogue as available on 2 October 2020.

The second goodness of fit is derived in a similar fashion to the first value, but instead for the corresponding S/N distributions.

The third goodness of fit is a weighting factor, based on how well a simulation matches the observed FRB rate.
We start by calculating our rate as:
\begin{equation}
 r = \frac{n_{\text{frbs}}}{n_{\text{days}}}
\end{equation}
with rate $r$, number of detected FRBs $n_{\text{frbs}}$ and number of days $n_{\text{days}}$.
To avoid the simulation size affecting the results, we choose to normalise all rates by the rate obtained with the
\survey{htru} survey \citep{2010MNRAS.409..619K} using the same inputs.
This turns our weighting function into
\begin{equation}
 w = \frac{1}{\left(  \frac{ r_{\text{survey}} }{  r_{\text{htru}} }\right)_{\text{sim}}
 - \left(  \frac{ r_{\text{survey}} }{  r_{\text{htru}} }\right)_{\text{real}} }
\end{equation}
with the weight $w$, survey rate $r_{\text{survey}}$ and HTRU rate $r_{\text{htru}}$.
Here rate ratios are determined both for simulated detections (${\text{sim}}$) and real detections (${\text{real}}$).
As the TNS catalogue does not include the length of time spent observing, we use rates available in literature to obtain a real rate per survey.
These values can be found in Table~\ref{tab:rates}.

\begin{table}
  \centering
  \caption{
  FRB detection rates for various surveys.
  The upper rows show derived detection rates from literature, with the lower rows showing expected rates on basis of \frbpoppy simulations.
  }
  \label{tab:rates}
\begin{tabular}{cccc}
 \hline\hline                                                                                                              \\[-9px]
                                                       & Survey                & \hspace{-2mm}Rate (day$^{-1}$)\hspace{-4mm} & Reference             \\[1px]
 \hline                                                                                                                    \\[-9px]
 \multirow{4}{*}{\rotatebox[origin=c]{90}{Literature}} & Parkes-HTRU           & 0.08              & \citet{champion2016}  \\
                                                       & CHIME-FRB             & 2                 & \citet{chawla2017}    \\
                                                       & ASKAP-Incoh          & 0.2               & Private Communication \\
                                                       & WSRT-Apertif          & 0.2               & \citet{lk+21}\hspace{-2mm}         \\[1px]
 \hline                                                                                                                    \\[-9px]
 \multirow{7}{*}{\rotatebox[origin=c]{90}{Simulated}}  & \survey{parkes-htru}  & 0.08              &                       \\
                                                       & \survey{wsrt-apertif}\hspace{-1mm} & 0.3              &                       \\
                                                       & \survey{fast-crafts}  & 0.2             &                       \\
                                                       & \survey{puma-full}    &$2 \times 10^2$&                       \\
                                                       & \survey{chord}        & $6 \times 10$   &                       \\
                                                       & \survey{ska1-low}     & 1.4            &                       \\
                                                       & \survey{ska1-mid}     & $2 \times 10$             &                       \\[1px]
 \hline
\end{tabular}
\end{table}

To limit compute requirements we choose to model FRBs out to a maximum redshift of $z_{\text{max}}=1$.
Almost all FRBs in our initial observed sample have an implied redshift smaller than 1.
For non-localised FRBs, the redshift is only inferred from the measured DM.
To ensure we compare only measured parameters, not inferred ones, we limit our evaluation of both real and simulated populations to bursts with a DM$_{\text{tot}} \leq 950$, not redshift $z_{\text{max}}=1$.
That way, any selections effects apply equally to both sets.
Only 4 real FRBs were detected above the threshold, and were cut. We discuss the implications of this choice in Sect.~\ref{sec:discussion:mc}.

For each intrinsic population we simulate the surveying of several surveys.
We choose to use four surveys to constrain the intrinsic FRB population - Parkes-HTRU, CHIME-FRB, ASKAP-Incoh and WSRT-Apertif.
These four surveys cover most of the one-off FRB detections to date \citep[see the Transient Name Server;][]{tns}, and so provide a solid basis from which to establish the properties of the intrinsic FRB population.

\begin{figure*}
\centering
\begin{subfigure}{.45\textwidth}
  \centering
  \includegraphics[width=\textwidth]{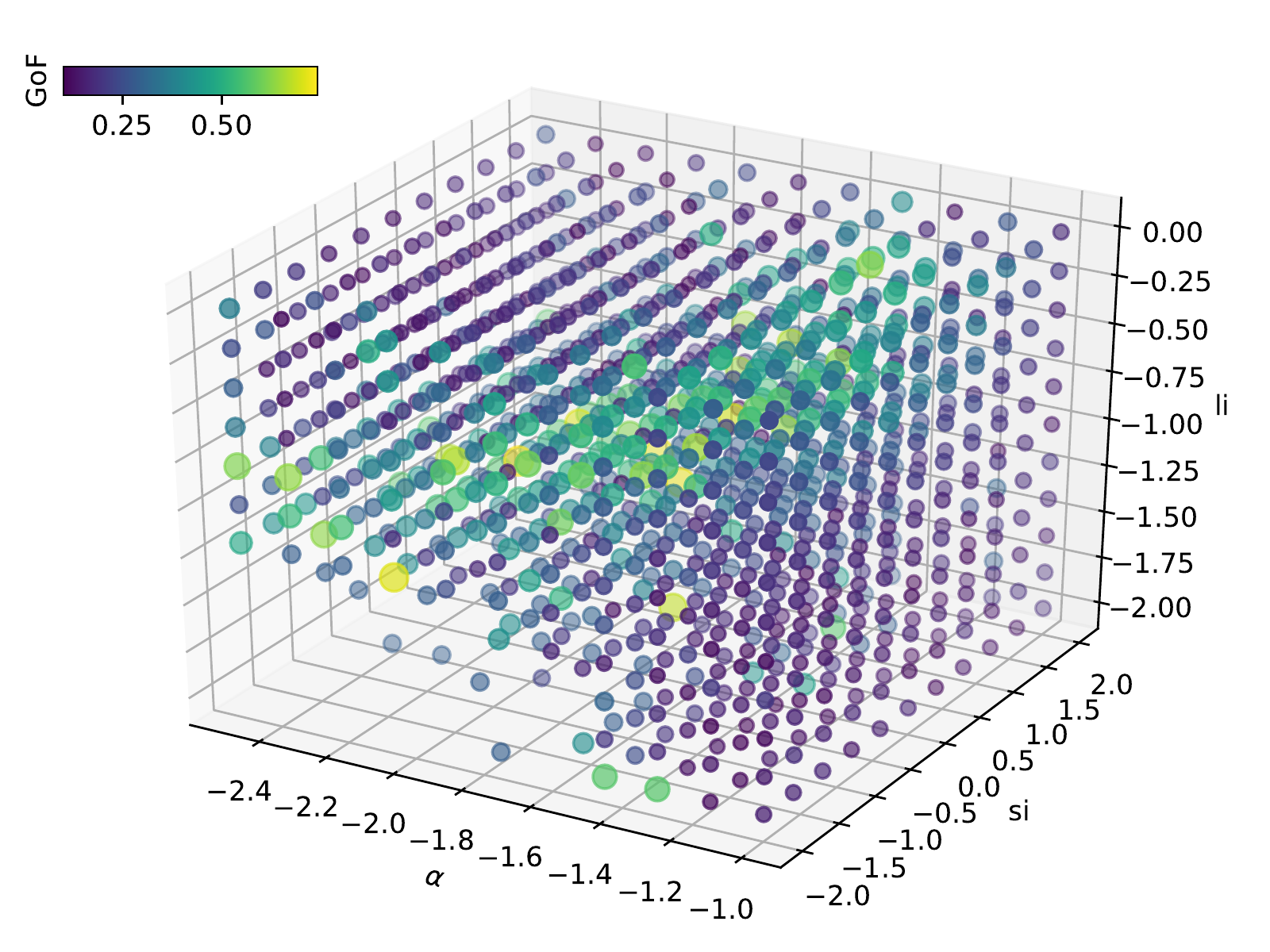}
\end{subfigure}%
\begin{subfigure}{.45\textwidth}
  \centering
  \includegraphics[width=\textwidth]{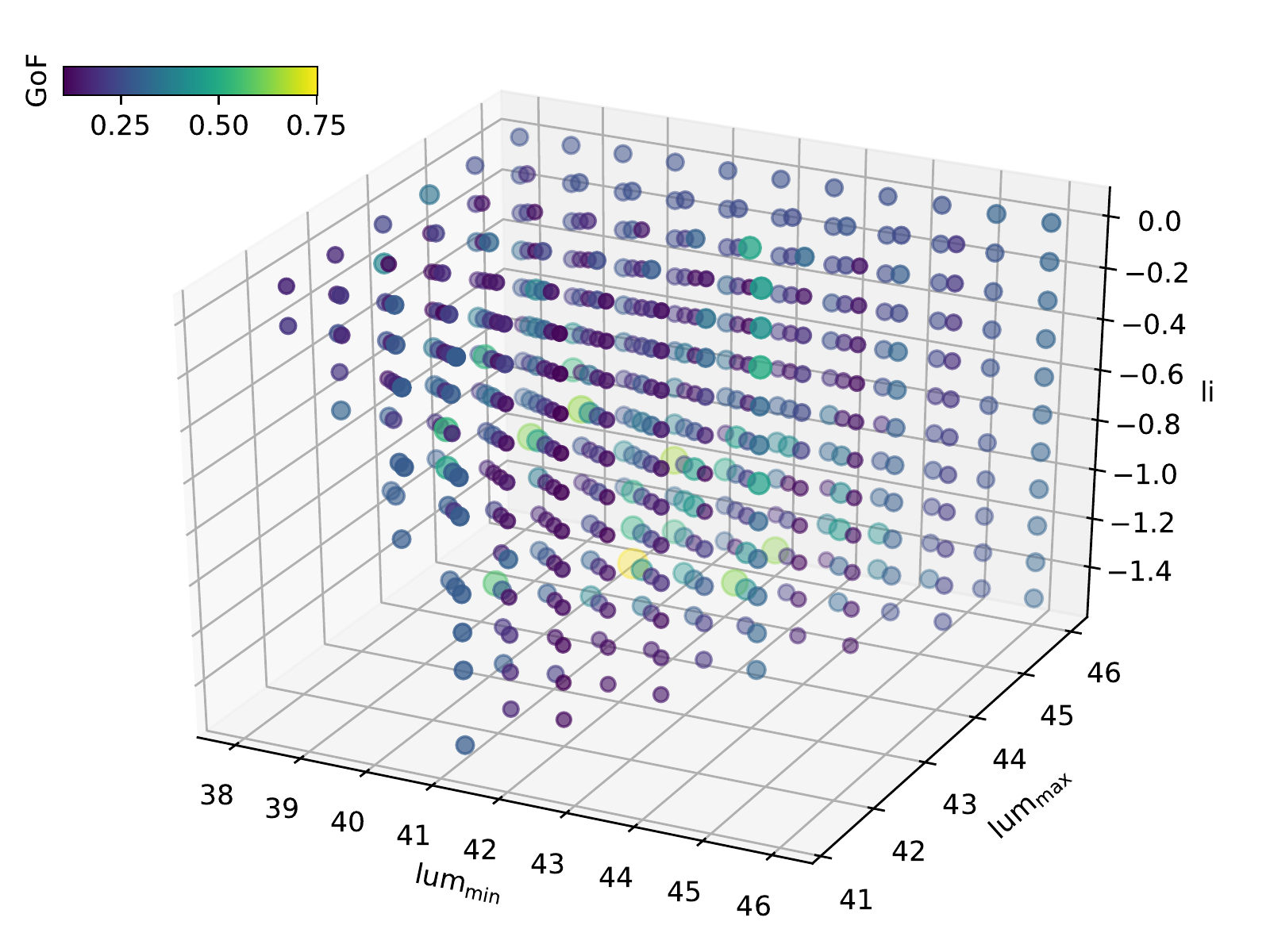}
\end{subfigure}
\caption{The Goodness-of-Fit (GoF) spaces spanned by two different sets of parameters. Higher GoFs are denoted both in brighter colours and larger markers. In the left panel an example is seen of a clear optimal region of a parameter space. This in comparison to the right panel, showing a parameter space with no clear optimal region. Each marker represents the weighted median GoF reflecting the results of multiple surveys in multiple dimensions.}
\label{fig:par_space}
\end{figure*}

Our aim is to derive an optimum set of values for each set of intrinsic population parameters being evaluated.
We start by generating populations with \pop{complex} inputs, shown in \citet{2019A&A...632A.125G} to be able to replicate Parkes-HTRU and ASKAP-Fly rates.
We then calculate a single, global GoF for each combination of values being evaluated
by taking the weighted median of all GoFs of populations modelled with the same input.
We choose to use this median as it damps the effect of outliers.
By including weights we additionally allow the GoF to reflect how well relative rates were simulated.
Within this newly constructed GoF space, the inputs that produce the highest GoF are marked optimum.
To gain an understanding of the underlying space,
we visually inspect sample runs.
This allows us to, e.g., avoid adding noise by including mostly featureless parameter spaces.
An example of this can be seen in Fig.~\ref{fig:par_space}, in which two GoF spaces over different parameters are visualized.
Most parameter spaces contain regions with clearly elevated GoF values (as seen in the left panel of
Fig.~\ref{fig:par_space}). For those, we use the optimum GoF values for the subsequent run.
In contrast, other parameter spaces, such as the one
seen in the right panel of Fig.~\ref{fig:par_space}
are quite  featureless.
These teach us the underlying parameter have little direct influence on the observed population.
We only evaluate these spaces in the first run. We found that when including such spaces
in subsequent runs, the somewhat random location of the exact optima added significant noise to the results.
After running through several parameter sets, we restarted the cycle until we convergence on a global optimum.

\section{Results}
\label{sec:results}
In this section, we present our simulations and outcomes. We discuss their implications in Sect.~\ref{sec:discussion}.

\subsection{LogN-LogS}
One of the main questions driving the field of FRBs is: what creates these bursts?
The answer requires an understanding of the conditions in which an FRB can be created.
This requires knowledge of the properties of the underlying source population.
With hundreds of catalogued FRB detections \citep{tns} it has become possible to probe this intrinsic source population.
Doing so is, however, challenging.
Firstly as selection effects prohibit a direct one-to-one mapping between observed and intrinsic distributions \citep[see e.g.][]{2019MNRAS.487.5753C}.
Secondly as observed distributions are often a single dimensional representation of a higher dimensional subspace of an intrinsic source population.

With \frbpoppy we can show how various intrinsic population properties can affect an observed distribution.
Such simulations also illustrate the difficulties in reversing the process to determine intrinsic population properties from an observed distribution.
An observed \lognlogs is a tantalizing distribution from which to infer population properties.
Such reasoning is tempting, as a deviation from a $-3/2$ slope expected from Euclidean universe can for instance be ascribed to non-Euclidean effects on source counts in $\Lambda$CDM.
The flattening of a \lognlogs distribution could indeed prove an invaluable and unique cosmological probe.
Nonetheless, we advise caution in such an interpretation as other intrinsic parameters can provide a similar effect, as
shown below.

Fig.~\ref{fig:flattening} displays the effects various intrinsic parameters have on  observed \lognlogs distributions.
Four panels are presented in this figure, each representing a different intrinsic parameter space.
Within each panel, we show the effect of varying the intrinsic distribution type on the observed distribution.
Differing line styles denote the cosmological effects on these variations, ensuring that both a simple, local, Euclidean population and populations out to high redshifts are represented.
The details of these populations can be found in Sect.~\ref{sec:methods:lognlogs}.
To help guide the eye, we plot $-3/2$ slopes with thin grey lines as reference.
As all simulations were run with a \survey{perfect} survey, with per definition an arbitrary S/N threshold, we choose to normalise all S/N values to a minimum value of 1 (i.e.~left-align the distributions), allowing for a clean comparison between various trends.

The diverse effects seen in Fig.~\ref{fig:flattening} serve as an aid for those trying to couple theory to observational expectations and the inverse.
For us, the similarity between distributions shows the clear need for a multi-dimensional approach to properly investigate the intrinsic FRB source population.

\begin{figure}
 \centering
 \includegraphics[width=\hsize]{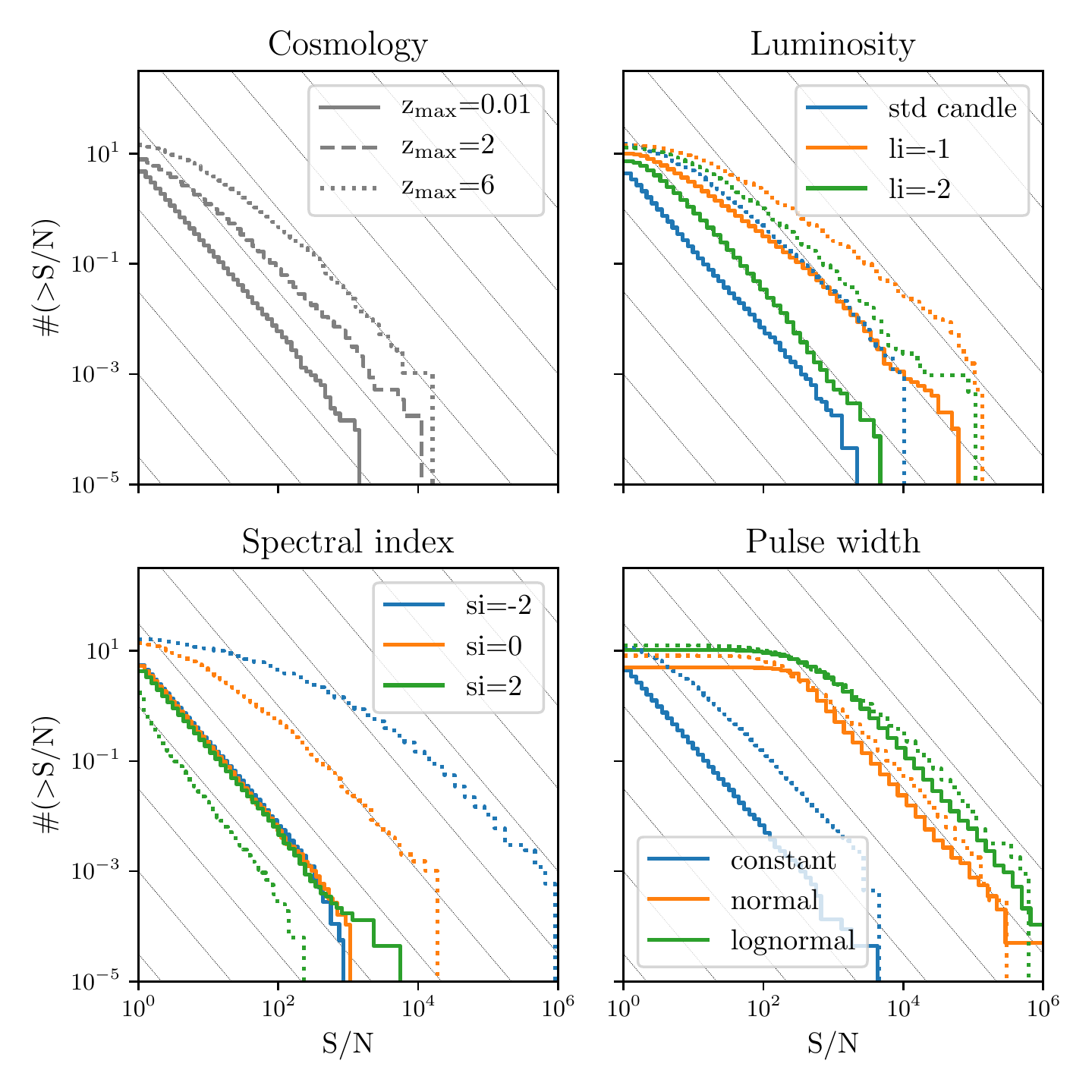}
 \caption{Cumulative distribution of the number of detected FRBs greater than a limiting signal-to-noise (S/N) ratio.
  Each panel shows the effect on the detected S/N ratios from three different inputs for the parameter denoted above the panel.
  Each panel additionally shows the cosmological effect on the various input distributions, denoted by the line styles given in the top left plot.
  Simulated observed S/N distributions can be seen for intrinsic source populations spanning various redshifts (top left), various intrinsic luminosity distributions (top right), various spectral indices (bottom left) and various intrinsic pulse width distributions (bottom right).
  Note that considering the arbitrary nature of a S/N threshold for a \survey{perfect} survey, all distributions have been normalised to a S/N ratio of one (i.e. the distributions have been left aligned).
  This allows for a cleaner comparison between trends.
  Each panel additionally features thin grey Euclidean lines with a slope of $-3/2$ such that various distributions can easily be compared.
 }
 \label{fig:flattening}
\end{figure}

\subsection{Monte Carlo}
Much remains unknown about the properties of the intrinsic FRB population \citep[see][]{petroff2019review, cordes2019review}.
Monte Carlo simulations allow us to build a coherent picture of this population,
by exploring how reasonable different areas of the source parameter space are.
Rather than trying to invert the observed parameter space, which is biased and incomplete,
population synthesis attempts to recreate the full underlying picture.
We look at many hypothetical underlying FRB populations through the lens of each survey,
and investigate which  view best  matches the actually detected burst set.

\begin{figure*}
 \centering
 \includegraphics[width=0.8\hsize]{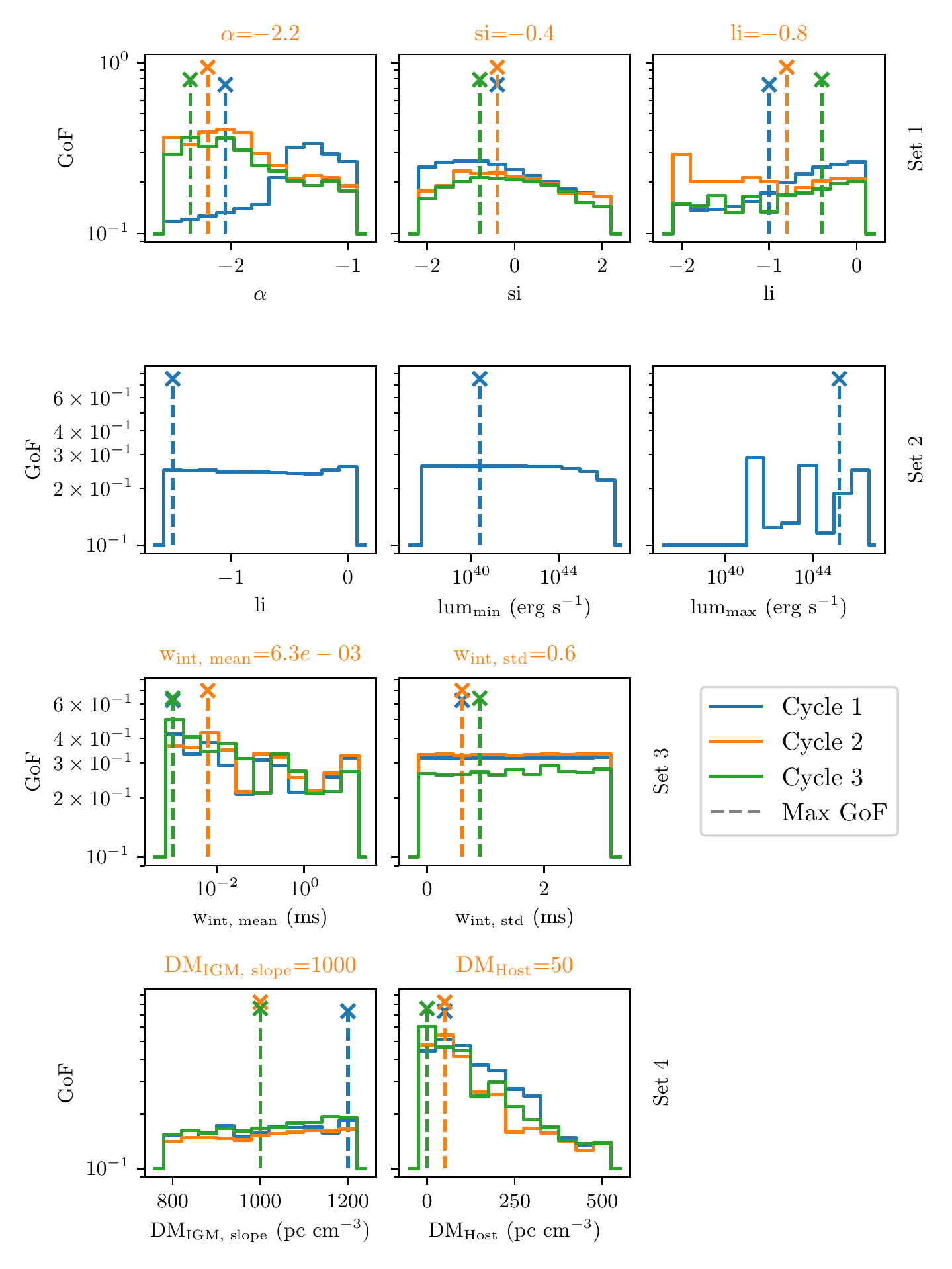}
 \caption{
 An overview showing how well various intrinsic parameter values can describe current one-off FRB observation, all expressed in a Goodness of Fit (GoF).
 Each row shows the set of parameters over which was iterated.
 Each panel shows the result of multiple runs, allowing the convergence to be checked.
 Each run shows the weighted median of GoFs for each parameter value, with a marker denoting the maximum GoF within that parameter space.
 From top left to bottom right, \lognlogs slope $\alpha$, spectral index $\text{si}$, luminosity index $\text{li}$, minimum luminosity $\text{lum}_{\text{min}}$, maximum luminosity $\text{lum}_{\text{max}}$, mean intrinsic pulse width $\text{w}_{\text{int, mean}}$, standard deviation intrinsic pulse width $\text{w}_{\text{int, std}}$, Macquart index $\text{DM}_{\text{IGM, slope}}$ and host dispersion measure $\text{DM}_{\text{Host}}$
 }
 \label{fig:mc}
\end{figure*}

In Fig.~\ref{fig:mc} we show the results of our Monte Carlo simulations.
We have split our global parameter space into subsets, and
 loop over these,
 to avoid the 60-day computational cost of directly searching the complete 11-dimensional space.
 Each panel individually shows the weighted median distribution of GoFs for each run.
These GoFs reflect how well our simulated observations match real observations of a variety of surveys.
Additionally, each run shows the location of the maximum GoF within that parameter space, indicating which parameter values show the best fit to reality.
The optimum values resulting in the best fit across all parameter sets are shown above each panel.

To ensure our fits approach a global rather than a local maximum, we run additional cycles over all parameters, the results of which are shown in a orange and green.
The higher GoF values of a runs indicate our Monte Carlo to be converging on a global maximum, with lower values a divergence.
This provides a way to check whether subsequent cycles are heading in the right direction.

Various trends are noticeable in these panels, with some showing flatter distributions than others.
We found the GoF parameter space of set 2, the luminosity parameters, to be fairly featureless, as visible in the right panel of Fig.~\ref{fig:par_space}.
Test runs showed that subsequent iterations moved the optimum around without statistically significant improvement to the outcome.
We thus used the output of set 1 in all subsequent runs, and avoided the parameter space for subsequent cycles.

Our \pop{complex} input model produces simulated FRBs out to redshift $z_{\text{max}}=1$ (cf. Sect.~\ref{sec:methods:mc}).
To ensure our simulation can also describe FRB detections at high DMs \citep[see e.g.][]{2018MNRAS.475.1427B},
that appear to be emitted farther out,
we additionally model the optimum population to a higher redshift of  $z_{\text{max}}=2$. We find this population is
still equally able to describe the observed DM and S/N distributions, and the rates.

We expand on the interpretation of the results for each parameter in Sec~\ref{sec:discussion:mc}, and how these compare
to predictions in literature.
Together these optimal parameter describe  a best-fit intrinsic population  (see \pop{optimum} in
Table~\ref{tab:populations}).
From this best model, expected FRB rates for future surveys can be derived.

\subsection{Future surveys}
\begin{figure*}
 \centering
 \includegraphics[width=\hsize]{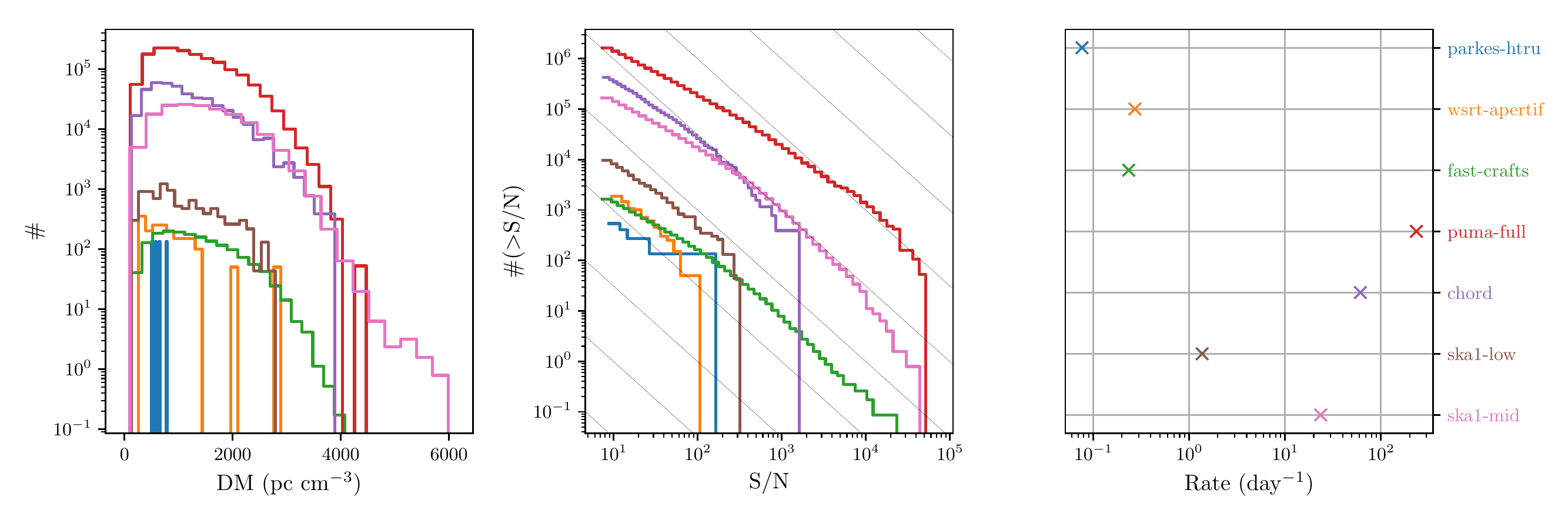}
 \caption{
  Expected observations of one-off FRBs by current and future surveys.
  \emph{left} Simulated dispersion measure distributions for various surveys.
  \emph{middle} \lognlogs distributions showing the number of detections above a S/N threshold for several surveys.
  \emph{right} Expected rates for each survey based on \frbpoppy simulations. All rates have been set such that the simulated Parkes-HTRU rate corresponds to the observed Parkes-HTRU rate.
  (We note that the size of this set of simulated detected Parkes-HTRU FRBs was large enough for producing reliable
  rates, but not large enough to properly sample the Parkes-HTRU distributions in the left panel.)  }
 \label{fig:future}
\end{figure*}

A large number of new radio observatories are in the construction or design phase: from
FAST-CRAFTS \citep{2019SCPMA..6259506Z}, to SKA-Low and
SKA-Mid \citep{dewdney2013ska1}, and e.g. PUMA-Full \citep{2020arXiv200205072C} or CHORD \citep{2019clrp.2020...28V}.
We simulate the detections expected for these surveys in comparison to current or past surveys such as Parkes-HTRU or WSRT-Apertif.
We derive these results using the \pop{optimal} population given in Table~\ref{tab:populations} together with the surveys presented in Table~\ref{tab:surveys}.
We adopt an Airy beam pattern for all of the surveys, as few have beam pattern information at this stage.
This allows for a cleaner comparison between the surveys.
Fig.~\ref{fig:future} displays the dispersion measure and \lognlogs distributions expected for these surveys.
The associated rates are displayed in the same figure, and listed in Table~\ref{tab:rates}.
Based on these numbers, we discuss the relative advantages of each survey in Sect.~\ref{sec:discussion:futuresurveys}.

\section{Discussion}
\label{sec:discussion}

In the previous section we reported the distributions and parameters that resulted from our simulations.
We here discuss with which source model these fit best.
We compare our outcomes against predictions by three source models.
First, the radio-loud magnetar models, where the bursts are generated as magnetically powered radio flares
\citep[e.g.][Model I]{2018MNRAS.477.2470L}.
Second,  spin-down powered models in which bursts are generated as supergiant pulses from young pulsars
\citep[e.g.][Model II]{2016MNRAS.457..232C}.
And third, models in which bursts are generated through blast wave masers \citep[e.g.][Model III]{2014MNRAS.442L...9L}.
For these three models classes, some quantitative expectations for FRB distributions exist.
These follow either from the observed behaviour of similar lower-powered systems,  or from theoretical considerations.
A number of  highly interesting other models, such as those involving
pulsar magnetospheres  swept back by strong plasma streams \citep{2018ApJ...854L..21Z} or
binary neutron-star mergers \citep{2013PASJ...65L..12T}, do not yet make predictions we can test against.

\subsection{LogN-LogS}
\label{sec:discussion:lognlogs}
In Fig.~\ref{fig:flattening} we show the effect of various intrinsic parameters on observed S/N distributions.

The resulting redshifted distributions seen in the top left panel of Fig.~\ref{fig:flattening} are understandable when considering effects arising over cosmological distances.
For a limiting redshift of 0.01, only the local universe is probed, resulting in a Euclidean universe where volume grows with $V \propto R^3$ over radius $R$.
As luminosity $L \propto R^{-2}$, the expected cumulative S/N distribution is expected to follow a slope of $-3/2$.
This is indeed observed in our simulations.
For higher maximum redshifts,  cosmological effects in which volume increases with less than $R^3$ start to play a role,
leading to a flattening of the \lognlogs distribution on the dim end \citep[see also][]{2019A&A...632A.125G}.
Additionally, time dilation reduces the relative observed rate of high redshift sources, further reducing the detection
rates of distant, and therefore faint, sources.
This too is seen in our simulations.

The top right panel of Fig.~\ref{fig:flattening} next shows the effect of various intrinsic luminosity functions.
As seen, a local population consisting of standard candles leads to a Euclidean \lognlogs distribution (a S/N distribution following a slope of $-3/2$).
Adopting a power law with a negative slope results in flattening of the distribution at low S/N bins.
This effect arises when sources have a limiting distance, yet a luminosity which could be detected further out were such sources also present at high redshifts.
This effect could also emerge upon beyond the cusp of a underlying number density distribution.
Due to this effect, we start to sample the underlying luminosity distribution.
The power law index has a strong effect on the resulting S/N distribution.
The steeper the slope, the more weight is placed at the lower end of the power law until it ultimately functions like a standard candle.
For this reason, the power law with a luminosity index of $-2$ shows a closer trend to that of standard candles, than the distribution with a luminosity index of $-1$.
Research that prefers  a steep luminosity index will accordingly need to take the distribution around the lower boundary of the power law into account for any predictions.
Indeed \citet{2018MNRAS.481.2320L} and \citet{2018ApJ...863..132F} advocate for a Schechter function, which partly addresses this problem.
The similarity between the local and cosmological \lognlogs distributions illustrates the difficulties in disentangling the effects arising from an intrinsic luminosity power law function and those from cosmology.
This is where the Macquart relationship, the link between redshift $z$ and intergalactic dispersion measure $\text{DM}_{\text{IGM}}$, can be of help \citep{2020Natur.581..391M}.

The bottom left panel of Fig.~\ref{fig:flattening} shows the effect of different spectral indices on observed S/N ratios.
With FRBs initially detected at 1.4\,GHz, the question quickly arose whether the same rate could be expected at different other observing frequencies.
Comparing the cosmological to local populations shows that these effects are only expected to be seen for distant sources.
While here we simulate detections with a \survey{perfect} telescope featuring a wide bandpass, for telescopes with a limited bandpass, a shift in spectral index could even affect the detection rate of local bursts.

The bottom right panel of Fig.~\ref{fig:flattening} shows the effect various pulse width distributions have on the observed \lognlogs.
Shifting from a constant pulse width to a wider distribution, such as a normal or log-normal distribution shows how bursts get smeared out over a wider range of S/N ratios.
For non-\survey{perfect} surveys, the sampling time will cause additional effects on the lower end of the \lognlogs distribution.
The presence of sources at high redshifts causes some bursts to redshift out of the observed time scale, resulting in a flattening of the \lognlogs.
Bursts at high redshifts, corresponding to the faintest bursts, are more heavily affected by this, hence the steady flattening of the \lognlogs towards lower S/N ratios.

These results show the necessity of an approach in which multiple aspects of an intrinsic source population are considered at the same time.
Monte Carlo simulations and population synthesis are an ideal approach to this problem.

\subsection{Monte Carlo}
\label{sec:discussion:mc}
\subsubsection{Best values}
Constraining the intrinsic properties of the FRB source population provides a method in which the origin of FRBs can be probed.
As evidenced from Fig.~\ref{fig:flattening}, only a multi-dimensional approach is feasible in disentangling the effects of multiple intrinsic distributions.
Our attempt at conducting such a multi-dimensional approach can be seen in Fig.~\ref{fig:mc}.
In the following paragraphs we discuss the results for each input parameter.

\paragraph{$\alpha$}
The variable $\alpha$ parametrizes our number density distribution.
This parameter is tied to the expected \lognlogs slope to provide for an easy interpretation, but is an intrinsic rather than an observed parameter.
On basis of volume scaling with $R^{3}$ and flux scaling with $R^{-2}$, one could expect that in ideal, non-cosmological circumstances $\alpha=-3/2$.
Indeed, as Fig.~\ref{fig:flattening} shows, this is the value to which the \lognlogs converges in the limit of the local universe.
Steeper values such as $\alpha=-2$ indicate not just a higher density of sources further out than close by, but in extension indicate an evolution in source density and thereby an evolution of progenitors.
On basis of our simulations we find $\alpha=-2.2$.
This is in line with results presented in \citet{2019MNRAS.483.1342J}, who argue for $\alpha=-2.2$ on basis of results
from ASKAP-CRAFTS, and with \citet{2018Natur.562..386S}, who argue for $\alpha=-2.1$.
In qualitative terms, this value for $\alpha$ fits with all three models outlined at the start of this section.
We expect the number density of neutron stars to be intrinsically related to the Stellar Formation Rate.
The prevalence of these neutron stars next determines the number density evolution of the magnetars required for both the hyperflare (Model I) and blast-wave (Model III) scenarios; but equally determines how many young pulsars exist, that could emit supergiant pulses (Model II).
This neutron-star density would then display an evolution in time.
Our result of $\alpha=-2.2$ for the FRB number density suggests such an evolution.
Our simulations are thus consistent with FRBs emerging from a cosmic population of magnetars or pulsars.

\paragraph{$\text{si}$}
The spectral index $\text{si}$ parametrises the relative peak flux densities of FRBs at different frequencies
(cf. Eq.~20 in \citealt{2019A&A...632A.125G}).
A negative value of $\text{si}$, for example,
indicates FRBs are brighter at lower frequencies \citep[for one such example, see][]{2020arXiv201208348P}.
If all other things are equal, more FRBs are then detected at 800\,MHz than at 1.4\,GHz.
In-band fits to ASKAP bursts at 1.4\,GHz produce a value of $\text{si}=-1.5^{+0.2}_{-0.3}$ \citep{2019ApJ...872L..19M}.
The \pop{complex} model with which we seed our population synthesis uses a similar value.
Fig.~\ref{fig:mc} shows that the histogram for $\text{si}$ alone initially peaks between  $-1.0$ and $-1.5$.
Models with that spectral index have, however, a poorer global goodness of fit when  $\alpha$ and $\text{li}$ are
included in the fit. For this ensemble of parameters, our simulations find  $\text{si}=-0.4$ after three cycles.
 This value is constrained predominantly by the relative detection rates of CHIME-FRB versus the three 1.4\,GHz
 surveys \citep[see also][]{chawla2017}.
The spectral index between bands thus does not agree with the in-band determination at 1.4\,GHz of \citet{2019ApJ...872L..19M}.
We conclude $\text{si}$ can not be modelled as a single power law over all radio frequencies \citep[also see][]{2019MNRAS.488.2989F}.
While at 1.4\,GHz FRB spectral behaviour may be similar to the $-1.4$ found in the galactic pulsar
population \citep{2013MNRAS.431.1352B},
the overall best value is significantly flatter.
Similar flat behaviour is seen in some FRB repeaters \citep[see e.g.][]{2019ApJ...876L..23H}.

Such flat spectral indexes are also found observationally in radio-loud magnetars (Model I).
The Galactic Center magnetar SGR J1745$-$2900 emits with a spectral index of $-0.4 \pm 0.1$ \citep{2015MNRAS.451L..50T},
similar to our findings for the FRB population.
Furthermore, magnetar spectra often can not be fit by a single power law \citep[e.g. PSR J1550$-$5418; ][]{Camilo_2008}.
Thus, spectral behaviour as seen in radio-loud magnetars fits our best models for  $\text{si}$.

If FRBs are generated as synchrotron masers in magnetized shocks produced by e.g. flares in
magnetars (Model III), the expected spectral indexes of the fluences are generally positive, not negative.
The four models for which \citet{2019MNRAS.485.4091M} present theoretical FRB light curves,
have an average spectral index of the fluence around 1.4\,GHz of $+0.5 \pm 0.5$.
This positive spectral index does not agree with the outcome of our simulations.

To evaluate if a supergiant-pulse model could also produce flat spectral indexes,
we extrapolate from the Crab giant pulses.
Those exhibit a steep, not flat, $\text{si}=-2.6$ index around 1\,GHz (although they do flatten,
to $\text{si}=-0.7$,  at the very low frequencies around 0.1\,GHz;
 \citealt{2017ApJ...851...20M}).
While the Crab giant pulses thus also show the power-law break we require,
 the values for $\text{si}$ are significantly higher than our models allow.
 Based on this very limited sample of one source we infer a supergiant-pulse origin is less likely.

\paragraph{$\text{li}$}
The luminosity index ${ \text{li} }$ parametrises an intrinsic luminosity function
such that $N(L)\propto L^{{ \text{li} }}$.
This differs from other notation styles in which $N(>L)\propto L^{x}$ or $dN/dL\propto L^{x}$, but allows for an easier interpretation as $\text{li}=0$ corresponds to a situation in which all luminosities are equally likely.
These indices are interchangeable using $\text{li} = x + 1$, though care must be taken whether $x=|x|$.
In \citet{2019A&A...632A.125G} the parameter ${ \text{li} }$ was labelled L$_{\rm bol,\ index}$.
While plotting fluence over excess dispersion measure shows that FRBs can not be described by standard candles \citep{petroff2019review}, the spread in intrinsic FRB luminosities is unknown.
Results from both \citet{luo2020luminosity} and \citet{2020MNRAS.tmp.3335Z} suggest a value of ${ \text{li} }=-0.8$.
Note this value has been converted to the definition of luminosity index used throughout this paper.

Our derived value of ${ \text{li} }=-0.8$ recovers this value in an fully independent fashion, suggesting a true constraint on the luminosity function is possible.

As we here present results on the population of one-off FRBs,
the luminosity index ${ \text{li} }$ we find describes how brightly each FRB
emits, once.
It is thus interesting to compare or contrast our value with
the luminosity index on the individual bursts of individual repeating FRBs.
A priori these values could be very different, especially if one-off and repeating sources are unrelated.
The burst distribution of FRB~121102, for example, was initially described with a powerlaw index of $-$0.8
\citep{law17}; while later analysis produced a steeper slope, of $-$1.7 \citep{2020A&A...635A..61O}.
If the former is correct, our best-fit brightness of bursts for one-off FRBs could potentially be drawn from the pulse
distribution of repeaters. We thus find that the intrinsic variation in repeating FRBs can explain the variation seen
between one-off FRBs.

Which sources could physically power the FRB population, and produce these power law brightness distribution?

  Our best-fit index is significantly flatter then generally seen in the giant pulses of radio pulsars (cf.~Model II).
In the Crab pulsar, one of the best studied giant-pulse emitters, the measured indices range from about
$-$1.3 to $-$2.0 \citep{Bhat-2008,2020A&A...634A...3V}.

Radio-loud magnetars (Model I) fit better.
As these sources are rare and not often active, few pulse distributions are available in the literature.
For XTE~J1810$-$197 enough statistics were accumulated though; it  emits radio bursts  with peak fluxes that follow a
powerlaw with index $-0.95 \pm 0.30$ \citep{Maan19}. That is in agreement with the FRB ${ \text{li} }$
we find.

Simulations of the blast-wave maser scenario (Model III) by \citet{2019MNRAS.485.4091M} find that the resulting FRB fluence is a function of the input flare energy.  To estimate the energy distribution of these flares, we look at the fluence distribution of magnetar bursts.
SGR~J1550$-$5418, for example, follows a slope of $-0.7 \pm 0.2$ at the high-energy end \citep{van_der_Horst_2012}.
The x-ray burst fluence distributions of other magnetars follow similar power laws.
We conclude our outcome value for the luminosity index ${ \text{li} }$ is generally consistent with Model  III.

\paragraph{$\text{l}_{\text{min, max}}$} The boundary values of a luminosity function could provide constraints on the emission mechanism of FRBs.
The flat nature of the parameter space seen in our Monte Carlo simulation  indicates the range of values
that is allowed is far wider
than our choice of model input.
This result is in contrast to results from \citet{2020ApJ...891...82W} arguing for a narrow energy band, but is line with with the recent detections of FRB-like bursts originating from the magnetar SGR 1935+2154 that span many orders of magnitude \citep[see e.g.][]{2020Natur.587...54T, 2020NatAs.tmp..232K}.
Extending this reasoning to cosmic FRBs would require an emission mechanism capable of producing bursts over an even
wider range than the  $10^{38} -10^{45}~\text{ergs s}^{-1}$ span we considered here.

\paragraph{$\text{w}_{\text{int}}$}
The intrinsic FRB pulse width distribution $\text{w}_{\text{int}}$ is a matter of considerable debate \citep[see, e.g.,][]{fonseca2020, 2020MNRAS.497.3076C}.
While from observations it is clear that the intrinsic distribution must cover millisecond values, selection effects
due to the instrumental response may shroud a large fraction of the population from us \citep[see][]{2019MNRAS.487.5753C}.
In our simulations we adopt a log normal distribution, where the values of $\text{w}_{\text{int, mean}}$ and $\text{w}_{\text{int, std}}$ indicate the desired mean and standard deviation of the distribution.
The goodness-of-fit in our simulations is relatively insensitive to the exact values of these two
parameters (Fig.~ \ref{fig:mc}, third row).
All models where  $\text{w}_{\text{int}}$ is relatively flat around 1\,ms, where the bulk of the detections occurs,
prove acceptable.

\paragraph{$\text{DM}_{\text{IGM, slope}}$}
The Macquart relation \citep{2020Natur.581..391M}, the DM-$z$ relationship can be expressed as $\text{DM}_{\text{IGM}} \simeq \text{DM}_{\text{IGM, slope}} z$ with $\text{DM}_{\text{IGM, slope}}$ the slope of this relationship.
Establishing the value of this parameter is crucial for the use for FRBs
for cosmological purposes such as establishing the baryonic distribution of the universe.
In \frbpoppy we model this relationship with a spread by applying a normal distribution $N$ to the DM of the intergalactic dispersion measure: $\text{DM}_{\text{IGM}} = N(\text{DM}_{\text{IGM, slope}}z, 0.2\,\text{DM}_{\text{IGM, slope}}z)$.
While a value of $\text{DM}_{\text{IGM, slope}}\simeq 1200\ \text{pc cm}^{-3}$ was commonly adopted on basis of \citet{ioka2003},
\citet{petroff2019review} argue for $\text{DM}_{\text{IGM, slope}}\simeq 1000\ \text{pc cm}^{-3}$ using \citet{2016ApJ...830L..31Y},
and \citet{cordes2019review} argue for $\text{DM}_{\text{IGM, slope}}\simeq 977\ \text{pc cm}^{-3}$.
Our finding of $\text{DM}_{\text{IGM, slope}}\simeq 1000\ \text{pc cm}^{-3}$ fits well within this expected band, though the fairly flat distribution suggest this to only be a weak constraint.
Adopting such a value for $\text{DM}_{\text{IGM, slope}}$ further suggests a higher contribution can be attributed to $\text{DM}_{\text{IGM}}$ than expected on basis of the single sight-line to FRB121102 \citep{2019ApJ...886..135P}.

\paragraph{$\text{DM}_{\text{Host}}$}
In \frbpoppy we use $\text{DM}_{\text{Host}}$ to reflect the combined DM contribution  of
the general host galaxy and any specific dense local environment in the host rest frame of the FRB source.
This avoids the challenging task of disentangling these contributions.
Our derived value of $\text{DM}_{\text{Host}}=50\ \text{pc cm}^{-3}$ is the same value commonly adopted by assuming the host galaxy to have properties similar to the average Milky Way DM contribution \citep[see e.g.][]{2018Natur.562..386S}.
Nonetheless, here too a large range of values have been forwarded in the literature.
\citet{cordes2019review} conclude on basis of Balmer-line estimates that host contributions could range from $\approx 100-200\ \text{pc cm}^{-3}$, where \citet{2020A&A...638A..37W} argue for a broad distribution of values centred around $50\ \text{pc cm}^{-3}$. and \citet{2017ApJ...839L..25Y} derive a far higher value of $270\ \text{pc cm}^{-3}$.
These results, however, have implied assumptions (such as a narrow luminosity distribution in \citealt{2017ApJ...839L..25Y}) that we do not make.
Our conclusion is based on a method that finds the overall best model, and does not only focus on DM.
The low contribution implies that FRBs either go off in low-mass galaxies, or at the outskirts of more massive ones.
This is corroborated by the relatively large offsets recently found for the localised FRBs in \citet{Macquart_2020}.
Such offsets do not immediately agree with source classes that closely follow stellar density and activity.
Young magnetars would certainly fall under that category \citep[e.g.][]{bochenek2020localized}.
The low DM values we favour offer no specific support for such a model, but would suggest it to be unlikely that FRBs emerge from supergiant pulses from young pulsars from which $\text{DM}_{\text{Host}}=10^2-10^4\ \text{pc cm}^{-3}$ would be expected \citep{2016MNRAS.458L..19C}.\\

We conclude that the optimal values that we derive are capable of describing the one-off FRB detections by
\survey{parkes-htru}, \survey{chime-frb}, \survey{wsrt-apertif} and \survey{askap-incoh}.
Our values, derived in an independent fashion, show good agreement with prior research into various aspects of the FRB population.
This shows the strength of a population synthesis, and provides a strong incentive to further constrain the intrinsic properties of the FRB population through population synthesis with a larger number of FRB detections.

We compared our values to three source models.
Model I, magnetically powered radio flares from magnetars, is generally consistent with the values we find.  Model II,
the  spin-down  powered supergiant pulses from young pulsars, cannot immediately account for the flat spectral indexes we
find.
Model III, masers from magnetar-flare shocks, predicts spectra that are inverted from our best fits.

We thus find
Model I fits best, and conclude that a cosmic population of magnetars producing radio flares
can be the source of  the observed Fast
Radio Burst sky.

\subsubsection{Limitations}
\label{sec:discussion:limitations}
In the interpretation of our results, some of the known limitations of our current Monte Carlo implementation should be
kept in mind. We made three choices, detailed below, that narrowed the scope of this first investigation, such that
we could run our simulations on a single powerful workstation (CPU: 12 cores, RAM: 128 GB). The results we present here serve as encouragement
for future investigations using \frbpoppy more massively parallel, on supercomputers. A fourth limitation is data availability.

Firstly, our simulations only concerned one-off FRB sources.
This has meant that  information provided by repeating sources, such as cadence and repeat rate, remains unused.
In principle the code is capable of including this time dimension \citep[as shown in][]{2020arXiv201202460G},
 but at prohibitive computational cost ($10^{n_{bursts}}$ times slower).
 Nonetheless, the conclusions drawn from our simulation should continue to hold for one-off sources,
 whether these emerge from the same population as repeaters or not.

Secondly, the derived optimum values are limited by the resolution of our simulation.
We choose to iterate over a maximum of three parameters at a time, to remain within a reasonable
compute wall time, of order weeks.
To limit this parameter space, each parameter was evaluated for its goodness-of-fit at just eleven points.
The resulting goodness-of-fit maximum within this parameter space could therefore only be establish at the intrinsic resolution of each parameter.
Furthermore, spanning the parameter subspace using this brute-force grid means much compute time is spent in areas with
low goodness-of-fit.
Modelling and understanding the surface of the goodness-of-fit parameter space,
using e.g. adaptive gridding, and numerical gradient ascent to find the optimum,
would allow not just for better values to be derived, but for faster optimization within the parameter space.

Thirdly, we chose to explore larger parameter spaces, over deriving errors on the first-found best values.
As a result of choosing to use our computational resources on a larger parameter space, we no longer have the resources to vary the input parameters to derive errors on the values.
Ideally one would run the same simulations multiple times, with different seed values, to estimate the error
values and contours.

Fourthly, our results are only as good as our modelling assumptions.
In \citet{2019A&A...632A.125G, 2020arXiv201202460G} we showed our modelling was able to replicate real one-off and repeater distributions, and in this paper we showed that the \pop{optimum} population is able to replicate detections from Parkes-HTRU, CHIME-FRB, ASKAP-Incoh and WSRT-Apertif.
Nonetheless, these results only hold for the observed distributions at that point in time.
Future observation may show an unknown selection effect or intrinsic parameter to be crucial in replicating observed distributions.

Taking these limitations into account, we next model the expected FRB detections for a range of future surveys.

\subsection{Future surveys}
\label{sec:discussion:futuresurveys}
An estimation of the expected FRB detection rates of various future FRB surveys  helps inform planning, commissioning and evaluation.
We populate a simulated universe with following the \pop{optimal} input parameter set (see Table~\ref{tab:populations}).
This fills a galactic volume up to a maximum redshift of $z_{\text{max}}=2.5$.
We next evaluate the simulated detections of future FRB surveys (see Fig.~\ref{fig:future}).
We scale our  simulated rates such that for Parkes-HTRU we  equal the real-work detection rate.
From this we determine the detection rate of future surveys, as given in Table~\ref{tab:rates}.
Using  realistic input population we do, however, produce different rates than previous estimates.
Where \citet{2019clrp.2020...28V}, for instance, expect CHORD to detect $\sim25$~FRBs per day, our simulations produce
$\sim60$~FRBs per day.
For PUMA,  \citet{2020arXiv200205072C} predict detecting 3500~FRBs per day;  we find a different value, of $\sim200$~FRBs per day.
One of the main reasons for these different outcomes is the value of $\alpha$.
A slight shift in the number density function can add significant weight to close-by or distant FRB sources, rapidly
changing the relative haul of deep versus wide surveys.
While our results are simulated to a maximum redshift $z_{\text{max}}=2.5$, extending the trend of DM distributions
through to higher values shows that surveys such as CHORD and PUMA might be capable of probing helium re-ionization at redshifts between 2--3 \citep[see e.g.][]{2018MNRAS.474.1900M, 2019MNRAS.485.2281C}.
Our results do not currently show whether FRBs could emerge from the era of hydrogen reionization, requiring bursts from redshift $z>7$ \citep{2018NatAs...2..865K}.
Nonetheless, the high detection rates predicted for future FRB surveys indicate enough high-redshift FRBs may be found to study potential dark matter halos and gravitational lenses.

\subsection{Opportunities, uses and future work}
\label{sec:discussion:next}
The open-source and modular nature of \frbpoppy aims to provide an easy-to-use tool for the FRB community.
All of the code used throughout this paper is available in \texttt{v2.1} of \frbpoppy, and is therefore available for use by others.

Further efforts towards deriving intrinsic FRB population properties through FRB population synthesis, with \frbpoppy or not, are strongly encouraged.

First, in the immediate future, the expected publication of hundreds of one-off FRBs from CHIME/FRB
will provide a wealth of input data for placing new, stronger constraints
on the intrinsic FRB population through modelling.

Next, now that we have demonstrated the validity of the Monte Carlo method in this paper,
extending and improving it could lead to substantial progress in the field.
A more parallelized supercomputing simulation will better determine the global maximum for the input population parameters.
Such an increase in compute resourcing (or efficiency) will also allow us to step beyond our current
limitation to only one-off FRB sources, and add repeating FRB sources \citep{2020arXiv201202460G}.
That would allow for further  insights into the number and type of  FRBs that inhabit the universe,  or even inform us
of the  composition of its possible  subpopulations.

\section{Conclusions}
\label{sec:conclusions}
We have constructed a Monte Carlo simulation capable of
producing a self-consistent underlying FRB  population that adequately recreates the sky as surveyed.
We thus derive the intrinsic properties of the one-off FRB population.
While the outcome from certain prior studies
matched our results, our new findings  were produced from a single and
coherent set of population parameters.
Our conclusions can be summarised as follows:
\begin{enumerate}
\item   Using a single observed parameter distribution (for example a \lognlogs distribution) to derive properties of the intrinsic FRB source population only provides weak constraints on underlying population parameters.
 A multi-dimensional approach is more informative.
 \item Through a Monte Carlo population synthesis, our \texttt{optimal} population is able to describe the DM and S/N
   distributions plus the rates of the one-off FRB detections of Parkes-HTRU, CHIME-FRB, WSRT-Apertif and ASKAP-Incoh.
 Our results are in strong agreement with prior studies.
\item Although the $\text{DM}_{\text{Host}}$ distribution dissents by a single order of magnitude, the spectral and luminosity index  of this \texttt{optimal} population, and its number density,
are consistent within the errors, with an FRB source population consisting of magnetars.
 \item Using this \texttt{optimal} population, we derive the expected rates and distributions for future FRB surveys.
 Our results indicate future FRB surveys will have high enough detection rates to use FRBs as cosmological probes.
\end{enumerate}
These conclusions demonstrate the value of FRB population synthesis in deriving the properties of the intrinsic one-off FRB population.

\begin{acknowledgements}
 We thank Liam Connor and Emily Petroff for early discussions.\vspace{1ex}\\

 The research leading to these results has received
 funding from the European Research Council under the European Union's Seventh Framework Programme (FP/2007-2013) / ERC
 Grant Agreement n. 617199 (`ALERT'); from Vici research programme `ARGO' with project number
 639.043.815, financed by the Netherlands Organisation for Scientific Research (NWO); and from
 the Netherlands Research School for Astronomy (NOVA4-ARTS).\vspace{1ex}\\

 We acknowledge the use of NASA’s Astrophysics Data System Bibliographic Services and the FRB data hosted on the Transient Name Server \citep{tns}.\vspace{1ex}\\

 This research has made use of \package{python3}{python} with \package{numpy}{numpy}, \package{scipy}{scipy}, \package{astropy}{astropy}, \package{pandas}{pandas}, \package{matplotlib}{matplotlib}, \package{bokeh}{bokeh}, \package{requests}{requests}, \package{sqlalchemy}{sqlalchemy}, \package{tqdm}{tqdm}, \package{joblib}{joblib}, \package{frbpoppy}{\paperone} and \package{frbcat}{pipfrbcat}.
\end{acknowledgements}

\bibliographystyle{aa}
\bibliography{bibliography}

\end{document}